\documentclass{article}
\usepackage{spconf,amsmath,graphicx,verbatim}
\usepackage{subcaption, comment, multirow, multicol, makecell, amssymb, bm, textcomp, kotex, arydshln}
\usepackage{adjustbox}
\usepackage{mathtools}
\usepackage{cite, url}
\usepackage{tabularx}
\usepackage{hyperref}
\usepackage{xcolor}
\hypersetup{
    colorlinks=true,
    linkcolor=purple,
    filecolor=magenta,      
    urlcolor=purple,
}
\newcolumntype{Y}{>{\centering\arraybackslash}X}

\usepackage[subtle]{savetrees}

\title{RawNeXt: Speaker verification system for variable-duration utterances with deep layer aggregation and extended dynamic scaling policies}
\name{Ju-ho Kim, Hye-jin Shim, Jungwoo Heo, and Ha-Jin Yu\sthanks{$^*$Corresponding author.}\thanks{This research was supported and funded by the Korean National Police Agency. [Project Name: Real-time speaker recognition via voiceprint analysis / Project Number: PR01-02-040-17]}}
\address{School of Computer Science, University of Seoul}

\begin{document}
\ninept
\maketitle
\begin{abstract}
Despite achieving satisfactory performance in speaker verification using deep neural networks, variable-duration utterances remain a challenge that threatens the robustness of systems. 
To deal with this issue, we propose a speaker verification system called \textit{RawNeXt} that can handle input raw waveforms of arbitrary length by employing the following two components: 
(1) A deep layer aggregation strategy enhances speaker information by iteratively and hierarchically aggregating features of various time scales and spectral channels output from blocks. 
(2) An extended dynamic scaling policy flexibly processes features according to the length of the utterance by selectively merging the activations of different resolution branches in each block. 
Owing to these two components, our proposed model can extract speaker embeddings rich in time-spectral information and operate dynamically on length variations. 
Experimental results on the VoxCeleb1 test set consisting of various duration utterances demonstrate that RawNeXt achieves state-of-the-art performance compared to the recently proposed systems. 
Our code and trained model weights are available at \url{https://github.com/wngh1187/RawNeXt}. 
\end{abstract}

\section{Introduction}
\label{sec:intro}

Speaker verification (SV) is the task of determining whether the identity of an anonymous voice matches the target speaker. 
In general, SV is performed as a series of processes: extracting fixed-dimensional utterance-level features from utterances and calculating the similarities between the features. 
Herein, the utterance-level features (\textit{i}.\textit{e}., speaker embeddings) are usually extracted through the network trained by the embedding learning methods. 
Due to advances in deep learning, deep neural network (DNN)-based embedding learning approaches such as d-vector \cite{variani2014deep} and x-vector \cite{snyder2018x} outperform traditional schemes (\textit{e}.\textit{g}., i-vector \cite{dehak2010front}). 
Although these methods have greater potential, they exhibit unsatisfactory performance for short input utterances \cite{zhang2017end}. 

In the field of SV, short utterances are one of the well-known performance degradation factors, increasing the uncertainty of embedding owing to insufficient speaker-specific information \cite{jung2019short}. 
To tackle this challenge, several studies have focused on how to effectively use the sparse speaker information contained in short utterances \cite{hajavi2019deep, gao2019improving}. 
They extracted speaker embeddings in a multi-scale aggregation (MSA) manner that adequately fuses and utilizes intermediate features of various time scales within the network. 
The MSA approach and its variants have shown superior performance for variable-duration utterance SV tasks \cite{gao2019improving, hajavi2019deep, jung2020improving}. 

\begin{figure*}[!t]
    \centering
    \includegraphics[width=\textwidth]{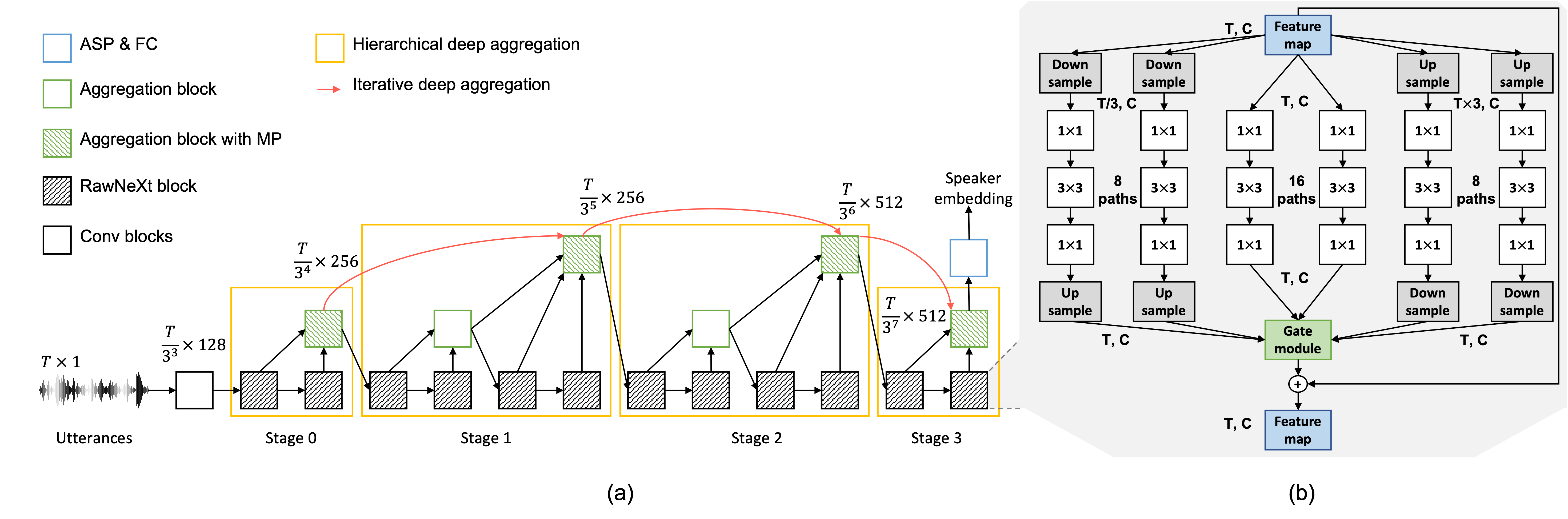}
    \caption{
    (a): Overall architecture of RawNeXt. 
    RawNeXt is trained to extract speaker embeddings rich in temporal and spectral information from utterances. 
    Iterative aggregation merges previous shallow stage to progressively propagate features of different resolutions. 
    Hierarchical aggregation combines different channels of blocks in stages to better refine the features. 
    (b): Structure of RawNeXt block. 
    The RawNeXt block divides the original paths in half and additionally processes the input into low- and high-resolution paths. 
    The features calculated for branches of each resolution are restored to their original resolution and aggregated by the gate module. 
    Through these processes, blocks can learn a dynamic scaling policy according to the input data. 
    }
    \label{fig:overall}
    \vspace{-0.2cm}
\end{figure*}

Inspired by the MSA to short utterances, we aim to advance the embedding extraction process by aggregating features in a more iterative and hierarchical fashion. 
To achieve this goal, we propose applying deep layer aggregation (DLA) \cite{Yu_2018_CVPR} as a speaker embedding extractor. 
DLA consists of two structures: iterative deep aggregation (IDA) and hierarchical deep aggregation (HDA). 
Like the MSA, IDA enriches the temporal information by merging features of different time scales from the previous stage (red lines in Fig. \ref{fig:overall} (a)). 
HDA fuses the channel axis of features from different blocks (yellow boxes in Fig. \ref{fig:overall} (a)). 
Typically, channels in a feature map ($\in \mathbb{R}^{T\times C}$) yielded from the 1d convolutional layer contain spectral information \cite{ravanelli2018sincnet}. 
Thus, HDA enhances the spectral information to extract more informative embeddings \cite{zhang2020deep}. 
Consequently, speaker embedding obtained by aggregating temporal and spectral information using DLA is expected to further improve the SV performance for variable-duration utterances. 

Meanwhile, the majority of SV systems process utterances in a fixed way with a series of manually designed layers \cite{jung2019short, hajavi2019deep, gao2019improving, jung2020improving}. 
However, it cannot be guaranteed to be optimal for variable-duration utterances. 
Therefore, the SV systems require dynamic speaker embedding extraction that can flexibly handle utterances of various lengths. 
Elastic \cite{wang2019elastic} was proposed for the scale variation of images, and it alleviated this issue by adding downsampling paths to the blocks. 
Elastic let the network dynamically process data by utilizing the appropriate resolution branches (original or downsampling paths) in each block. 
In this study, to process features according to the length of the utterance, we introduce an extended dynamic scaling policy (EDSP) based on the Elastic. 
Compared with the existing method, EDSP increases the resolution branch (upsampling paths) to provide more varied scaling options. 
In addition, the proposed method exploits a multi-head attention-based gate module to selectively aggregate the activation of each path. 
Thus, the EDSP encourages the model to better extract speaker information by dynamically responding to the length of utterances. 

In summary, DLA enriches the speaker's time-spectral information and EDSP induces flexible operation according to the length of input utterances. 
By applying both methods, we finally propose an SV system called \textit{RawNeXt} (suggesting the \textit{next} version of \cite{jung2020improved}), which is robust to length variation of utterances with raw waveforms as inputs. 
To train and evaluate the models, the VoxCeleb datasets \cite{nagrani2017voxceleb, chung2018voxceleb2} were used. 
As a result of the experiments, RawNeXt reported state-of-the-art performance for short-utterance SV tasks. 
Moreover, our proposed system showed results comparable to the top-three single systems of the 2020 VoxCeleb Speaker Recognition Challenge (VoxSRC-20) \cite{nagrani2020voxsrc}. 
Additionally, we demonstrated the effectiveness of RawNeXt's components through ablation and analysis experiments.

\section{Baseline}
\label{sec:baseline}

\begin{table}[!t]
 \caption{
 The architecture of the baseline. 
 The sample size of the input waveform is 59,049. 
 For convolutional layers, numbers inside parentheses refer to the filter length, stride size, and number of filters. 
 }
  \centering
  \label{tab:DNN_arch}
  \renewcommand{\tabcolsep}{1.2mm}
  \begin{tabular}{c|c|c|c}
  \Xhline{1pt}
  \textbf{Level}&\textbf{Block structure} & \textbf{\# Blocks } & \textbf{Output}\\
  \Xhline{1pt}
  \multirow{3}{*}{Convs} & Conv(3, 3, 128) & 1 & \multirow{3}{*}{2,187$\times$128}\\
  \cline{2-3}
   & Conv(3, 1, 128) & \multirow{2}{*}{2} & \\
    &  Maxpool(3) & \\
  \hline
  \multirow{2}{*}{Stage 0}& Conv(1,1,256) & \multirow{2}{*}{2} & \multirow{2}{*}{729$\times$256} \\
  & Conv(3,1,256), \textit{C}=32 & \\
  \cline{1-1}  \cline{3-4}
  \multirow{2}{*}{Stage 1}&  Conv(1,1,256) & \multirow{2}{*}{4} & \multirow{2}{*}{243$\times$256} \\
  \cdashline{2-2}
  & Maxpool(3) & \\
  \hline
  \multirow{2}{*}{Stage 2}& Conv(1,1,512) & \multirow{2}{*}{4} & \multirow{2}{*}{81$\times$512} \\
  & Conv(3,1,512), \textit{C}=32 & \\
  \cline{1-1}  \cline{3-4}
  \multirow{2}{*}{Stage 3}& Conv(1,1,512) & \multirow{2}{*}{2} & \multirow{2}{*}{27$\times$512} \\
  \cdashline{2-2}
  & Maxpool(3) & \\
  \hline
  Pooling & ASP & 1 & 1,024\\
  \hline
  Embedding & FC(512) & 1 & 512\\
  \Xhline{1pt}
  \end{tabular}
  \label{tab:baseline}
  \vspace{-0.2cm}
\end{table}

In SV research, novel approaches are emerging that deal directly with raw waveforms \cite{ravanelli2018sincnet, jung2020improved} rather than engineered acoustic features such as mel-filterbank energies \cite{variani2014deep, snyder2018x}. 
It is known that systems trained in a data-driven manner on less-processed data can extract discriminative representations suitable for SV tasks with minimal hyper-parameter search of acoustic feature pre-processing \cite{ravanelli2018sincnet, jung2020improved}. 
To take advantage of these pros, we use raw waveforms as model inputs. 

Table \ref{tab:DNN_arch} shows the baseline structure of this study based on ResNeXt \cite{xie2017aggregated}. 
The ResNeXt contains the grouped convolutional layers known as a split-transform-merge strategy in blocks and recently reported reliable performance in SV \cite{zhou2021resnext}.
Cardinality (C) refers to the number of groups in grouped convolution operations. 
Stages 0, 1 and 2, 3 have identical block structures, respectively, and max pooling (MP) is applied at the end of each stage. 
Batch normalization (BN) and ReLU activation are employed after every convolutional layer. 
Finally, speaker embeddings with 512 dimensions are extracted through the attentive statistical pooling (ASP) \cite{okabe2018attentive} and fully connected (FC) layers.

\begin{table*}[t!]
 \caption{Results of comparison with recently proposed speaker verification system for short utterances. ($^\top$: drawn from \cite{kye2021supervised}, $^{\dagger}$ : our implementation, *: data augmentation)
 }
    \label{tab:short}
    \centering
    \begin{tabular}{c c c c | c c c c}
        \Xhline{1pt} 
        \multirow{3}{*}{Model} & \multirow{2}{*}{Input} & \multirow{2}{*}{Loss} & \multirow{2}{*}{Aggregation} & \multicolumn{4}{c}{Vox1-O}  \\
        \cline{5-8}
        & \multirow{2}{*}{Feature} & \multirow{2}{*}{Function} & \multirow{2}{*}{Method} & EER\% & EER\% & EER\% & EER\% / $C_{det}^{min}$ \\
        & & & & 1s  & 2s & 5s & full\\
        \Xhline{1pt} 
        MSEA+FPM \cite{jung2020improving} & MFB-64 & A-Softmax & LDE & 5.92 & 3.38 & 2.17 & 1.98 / 0.205 \\
        ResNet34 \cite{kye2020meta}$^\top$ & MFB-40 & Softmax+PN & TAP  & 4.77 & 3 & 2.2 & 2.08 / 0.234 \\
        ResNet34 \cite{kye2021supervised} & MFB-40 & Softmax+PN & ANF  & 4.49 & 2.88 & 2.04 & 1.91 / 0.221 \\
        \hline{}
        RawNet2 \cite{jung2020improved}$^{\dagger}$ & Waveform & Softmax & ASP & 7.24 & 3.88 & 2.64 & 2.43 / 0.236 \\
        ResNeXt (Baseline) & Waveform & Softmax & ASP & 6.12 & 3.68 & 2.45 & 2.16 / 0.187 \\
        RawNeXt (Proposed) & Waveform & Softmax & ASP & \textbf{4.47} & \textbf{2.58} & \textbf{1.72} & \textbf{1.54} / \textbf{0.166}  \\
        \hline{}
        RawNeXt* & Waveform & AAM-Softmax & ASP  & \textbf{4.37} & \textbf{2.34} & \textbf{1.45} & \textbf{1.29} / \textbf{0.142} \\
        \Xhline{1pt} 
    \end{tabular}
    \vspace{-0.2cm}
\end{table*}

\section{Proposed methods}
\label{sec:proposed}

\subsection{Deep layer aggregation}
\label{subsec:dla}
Beyond constructing deeper and wider DNNs to increase accuracy, it is also being explored to connect blocks more closely \cite{lin2017feature}. 
By merging features of several layers, systems can yield context-rich representations for target tasks and mitigate the gradient vanishing problem by back-propagating earlier to lower layers \cite{lin2017feature, Tang2019DeepSE}. 
Furthermore, in the field of SV, it is known that the embeddings extracted from MSA-based models are robust to short-duration utterances \cite{hajavi2019deep, jung2020improving}. 
Similar to the direction of previous studies, we intend to derive speaker embeddings by fusing features in a more iterative and hierarchical manner for utterances of various lengths. 
To achieve this aim, we use the DLA \cite{Yu_2018_CVPR} as a speaker embedding extractor. 

Fig. \ref{fig:overall} (a) illustrates the overall architecture of the proposed system, RawNeXt. 
Compared with the baseline as in Table \ref{tab:baseline}, RawNeXt additionally utilizes IDA and HDA modules for feature aggregation at each stage. 
IDA iteratively merges stages, from shallow to deep. 
In this way, aggregations of different time resolutions enrich temporal context information in deep features. 
HDA hierarchically fuses blocks in a tree-structured fashion for each stage. 
Hence, the context information in spectral domain is enhanced by combining the feature channels of different levels.

The aggregation blocks learn to select important information from the multiple inputs and project it into a single output. 
The aggregation block, $N$, is formulated as follows: 
\begin{equation}
    N(x_1,...,x_n) = \sigma(Conv([x_1,...,x_n])))
\end{equation}
where $x_i$ is the output of the previous $i$-th block, and these outputs are concatenated, denoted as $[\cdot]$. 
After that, it is transformed into the single output via a convolution layer with a kernel and stride size of 1, followed by BN and ReLU activation functions denoted as $\sigma(\cdot)$. 
The last aggregation block of each stage applies MP to reduce the number of frames in the feature by one-third. 
Finally, the last feature output from Stage 3, containing the compressed network-wide information, is converted into speaker embedding through ASP and FC.

\subsection{Extended dynamic scaling policy}
\label{subsec:edsp}
The representations that do not consider the scale variation of the data are often sub-optimal for the target tasks \cite{9292935}. 
To mitigate this issue, Wang \textit{et al.} \cite{wang2019elastic} proposed a method, Elastic. 
This approach allows the network to learn a scaling policy from data, in which the system can decide among the original and the downsampling paths in each block. 
Thus, Elastic encourages the network to perform dynamically on the scale of data. 
Inspired by this scheme, we argue that the variable-duration SV task should also perform flexibly depending on the utterance lengths. 
Therefore, we propose an EDSP based on Elastic for arbitrary length utterances.

Fig. \ref{fig:overall} (b) shows the structure of RawNeXt block, in which the EDSP strategy is applied to the baseline block. 
The proposed EDSP reduces the original-resolution branches from 32 to 16 and extends the feature resolution range by adding eight downsampling and eight upsampling branches in parallel. 
Features are processed through convolutional layers of the same structure in each branch. 
Herein, applying the convolution with the same kernel and stride size in different resolutions implies extracting features with receptive fields of different sizes. 
That is, for the same input, the receive field sizes of the original path, downsampling path and upsampling path are three, nine, and one, respectively.
Therefore, the resolution path expansion at each block provides the versatility to process feature maps with a combination of various receptive fields compared to the fixed single-scale branches. 
Indeed, we observed that the required resolution of branches varies with the length of utterances (see Fig. \ref{fig:activation}). 
Subsequently, upsampling and downsampling are applied to the low- and high-resolution paths, respectively, to restore the original resolution. 
This process encourages features to be handled in multiple time scales by selectively activating each branch based on utterance lengths.

At low-, original-, and high-resolution branches, individually calculated features  $F^l(x), F^o(x),$ and $F^h(x) \in \mathbb{R}^{T\times C}$ are as follows: 
\begin{equation}
\begin{aligned}
    &F^l(x) = \sum_{i=1}^{8}U^l_{i}(f^l_{i}(D(x))),&\\
    &F^o(x) = \sum_{i=1}^{16}f^o_{i}(x),&\\
    &F^h(x) = \sum_{i=1}^{8}D(f^h_{i}(U^h_{i}(x)))& 
\end{aligned}
\end{equation}
where $f^r_i$ is the convolutional layer of the $i$-th path in the $r$ resolution branch, $r =\{l,o,h\}$. 
$D(x)$ is the downsampling function, which is an average pooling layer. 
$U^r_i(x)$ is the upsampling function, which is a transposed convolutional layer. 
Both have the same kernel and stride size of 3. 

Furthermore, we introduce a multi-head attention-based gate module to dynamically fuse the activation of branches. 
Firstly, the features output from the three branches are averaged based on the time axis and then concatenated, denoted by $\mu(\cdot)$ and $[\cdot]$, respectively. 
\begin{equation}
    H = [\mu(F^l(x)),\mu(F^o(x)),\mu(F^h(x))],\ H \in \mathbb{R}^{3 \times C}
\end{equation}
\begin{equation}
    W_t = Z^\top\sigma(Y^\top H_t + p) + q,\ W_t \in \mathbb{R}^{1 \times C}
\end{equation}
Afterward, the obtained vector $H_t$ is transformed to the attention weights $W_t$ through two linear layers ($Y,p$ and $Z,q$), where $t$ is the time axis. 
Then, the attention score, $A^c_t$ is derived by applying the softmax operation to each channel of $W$, where $c$ is the channel axis. 
\begin{equation}
    A^c_t = \frac{exp(W^c_t)}{\sum_{i=1}^{3}exp(W^c_i)}, \ A^c_t \in \mathbb{R}^{1}
\end{equation}

Consequently, the multi-head attention-based gate module can selectively reflect the activation of each branch by multiplication between the feature and attention map, $A_t \in \mathbb{R}^{1\times C}$ (at each multiplication term, the time axis of attention maps is broadcast). 

\begin{equation}
\begin{aligned}
\label{eq1}
    &Gate(F^l(x), F^o(x), F^h(x))&\\
    &\quad= F^l(x) \times A_1 + F^o(x) \times A_2 + F^h(x) \times A_3&
\end{aligned}
\end{equation}
Finally, RawNeXt block, $B$, is expressed as follows: 
\begin{equation} 
    B(x) = \sigma(Gate(F^l(x), F^o(x), F^h(x))+x)
\end{equation}
The stack of RawNeXt blocks increases the combination of resolution path options exponentially, leading to dynamic propagation for variable-length utterances.

\begin{table*}[t!]
 \caption{
 Comparison with VoxSRC-20's top-three single speaker verification systems \cite{nagrani2020voxsrc} on the three different evaluation trials. 
 }
    \label{tab:sota}
    \centering
    \begin{tabular}{c c c c | c c c}
        \Xhline{1pt} 
        \multirow{2}{*}{Model} & Input & Loss & Aggregation & Vox1-O & Vox1-E & Vox1-H  \\
         \cline{5-7}
        & Feature & Function & Method & EER\% / $C_{det}^{min}$ & EER\% / $C_{det}^{min}$  & EER\% / $C_{det}^{min}$ \\
        \Xhline{1pt} 
        ResNet-100m2 \cite{torgashov2020id} & MFB-80 & AM-Softmax & SP & 1.1 / \textbf{0.064} & - &  - \\
        DPN68 \cite{xiang2020xx205} & MFB-40 & CM-Softmax & SP & 0.77 / 0.077 & 0.96 / 0.103 &  1.66 / \textbf{0.156} \\
        ECAPA-TDNN \cite{thienpondt2021idlab} & MFCC-80 & AAM-Softmax & ASP & \textbf{0.56} / 0.074 & \textbf{0.84} / \textbf{0.096} &  \textbf{1.57} / 0.164 \\
        \hline{}
        RawNeXt & Waveform & AAM-Softmax & ASP & \textbf{1.29} / 0.142 & \textbf{1.17} / \textbf{0.138} & 2.28 / 0.236 \\
        RawNeXt & Waveform & AAM-Softmax+AP & ASP & 1.32 / \textbf{0.136} & 1.19 / 0.145 & \textbf{2.23} / \textbf{0.228} \\
        \Xhline{1pt} 
    \end{tabular}
    \vspace{-0.2cm}
\end{table*}

\begin{table}[!t]
 \caption{
 Ablation experiments of RawNeXt components. (D: Deep layer aggregation, E: Elastic, U: upsampling path, G: Gate module) 
 }
    \label{tab:ablation}
    \centering
    \resizebox{\linewidth}{!}{
    \begin{tabular}{l | c c c c | c c c c}
        \Xhline{1pt} 
        \multirow{2}{*}{Systems} & \multirow{2}{*}{D} & \multirow{2}{*}{E} & \multirow{2}{*}{G} & \multirow{2}{*}{U} & \multicolumn{4}{c}{Vox1-O (EER\%)} \\
         \cline{6-9}
        & & & & & 1s  & 2s & 5s & full\\
        \Xhline{1pt} 
        \#1(ResNeXt)&$\times$&$\times$&$\times$&$\times$& 6.12 & 3.68 & 2.45 & 2.16 \\
        \#2 &\checkmark&$\times$&$\times$&$\times$& 4.82 & 2.98 & 2.08 & 1.93 \\
        \#3 &$\times$&\checkmark&$\times$&$\times$& 5.39 & 3.18 & 2.16 & 1.95 \\
        \#4 &\checkmark&\checkmark&$\times$&$\times$& 4.66 & 2.94 & 2.13 & 1.94 \\
        \#5 &\checkmark&\checkmark&\checkmark&$\times$&
        4.67 & 3.01 & 2.08 & 1.88\\
        \#6 &\checkmark&\checkmark&$\times$&\checkmark& 
        4.65 & 2.81 & 1.94 & 1.82\\
        \#7(RawNeXt)&\checkmark&\checkmark&\checkmark&\checkmark&\textbf{4.47}& \textbf{2.58}&\textbf{1.72}&\textbf{1.54} \\
        \Xhline{1pt} 
    \end{tabular}
    }
    \vspace{-0.2cm}
\end{table}

\section{Experimental setup}
\label{sec:exp}

\subsection{Datasets}
\label{subsec:page}

For training, we used the VoxCeleb2 dataset \cite{chung2018voxceleb2}, which comprises over 1 million utterances from 6,112 speakers. 
To prove the effectiveness of our model under various conditions, we exploited three evaluation trials on the VoxCeleb1 dataset \cite{nagrani2017voxceleb}. 
The original evaluation trial (Vox1-O) consists of 37,611 enrollment-test utterance pairs from 40 speakers, corresponding to the test set of the VoxCeleb1 dataset. 
The extended evaluation trial (Vox1-E) contains a list of 579,818 pairs from 1,251 speakers in the entire VoxCeleb1 dataset, and the hard evaluation trial (Vox1-H) includes a list of 550,894 pairs with the same nationality and gender from 1,190 speakers. 
We tested trials using cosine similarity and evaluated the models with the equal error rate (EER) and the minimum detection cost function ($C_{det}^{min}$), as in \cite{chung2018voxceleb2}. 

\subsection{Implementation details}
\label{subsec:detail}

We employed the raw waveforms as input with pre-emphasis applied. 
For each iteration, the mini-batch consisted of 320 utterances (2 utterances from each of 160 randomly selected speakers). 
The lengths of the two utterances for each speaker were set to a fixed 59,049 samples and a random number of samples between 16,000 and 59,049. 
The random length utterances were duplicated to fit 59,049 samples for mini-batch construction. 
This configuration encourages the model to learn the EDSP strategy explicitly by training utterances of various lengths. 
In all experiments, we used the AMSGrad optimizer \cite{reddi2019convergence}. 
The initial learning rate (LR) was $1e^{-3}$ and decreased to $1e^{-7}$ for 80 epochs using a cosine LR scheduler. 
We set the weight decay to $1e^{-4}$. 
In several experiments, data augmentation was applied using room impulse response simulation and the MUSAN corpus \cite{snyder2015musan}.

\section{Results}
\label{sec:result}

In Table \ref{tab:short}, we compare our model with the recently proposed systems for short utterances. 
To evaluate the performance with short utterances on the Vox1-O trial, we used full-duration enroll utterances and test utterances truncated to durations of 1, 2, and 5 seconds. 
The test utterance was cropped in the middle of the utterance, and if the utterance length was shorter than the target length, it was duplicated. 
As a result of the experiments, baseline ResNeXt showed relatively satisfactory performance for the full-duration test, with a better result than RawNet2 under the same conditions. 
However, significant performance degradation occurred for the variable-duration scenario compared to the recently proposed systems (rows 1,2,3). 
Proposed RawNeXt outperformed other models with different input features or improved loss functions under all test conditions. 
In addition, RawNeXt with the combination of data augmentation and AAM-softmax \cite{deng2019arcface} loss achieved state-of-the-art results on short-utterance SV scenarios as well as full-duration utterances. 
Based on these results, we judge that the proposed model is effective for variable-duration utterances and can be enhanced by using various loss functions. 

Table \ref{tab:sota} shows the comparison with the top-three systems of VoxSRC-20, which reported state-of-the-art performance in SV. 
We trained RawNeXt using improved loss functions, such as AAM-softmax and angular prototypical network (AP) \cite{chung2020defence}, and all experiments in this table used data augmentation. 
Although our system was proposed for variable-duration utterances, it exhibited relatively tolerable performance compared to the top-three systems. 
These results suggest that RawNeXt has high potential for not only short utterances but also generalization of SV. 

Table \ref{tab:ablation} presents the results of ablation experiments to demonstrate the efficacy of each RawNeXt's component for variable-duration utterances. 
The results of Systems \#1, 2, 3, and 4 show that the use of DLA and Elastic, proposed for computer vision tasks, leads to better SV performance. 
This implies that the motivations of each method are well aligned with the goal of short-utterance SV. 
A comparison of Systems \#4, 5, 6, and 7 suggests that using both our proposed gate module and upsampling path (meaning EDSP) complements the original Elastic scheme, resulting in additional performance improvement for variable-duration utterances.

\begin{figure}[!t]
  \centering
  \includegraphics[width=\linewidth]{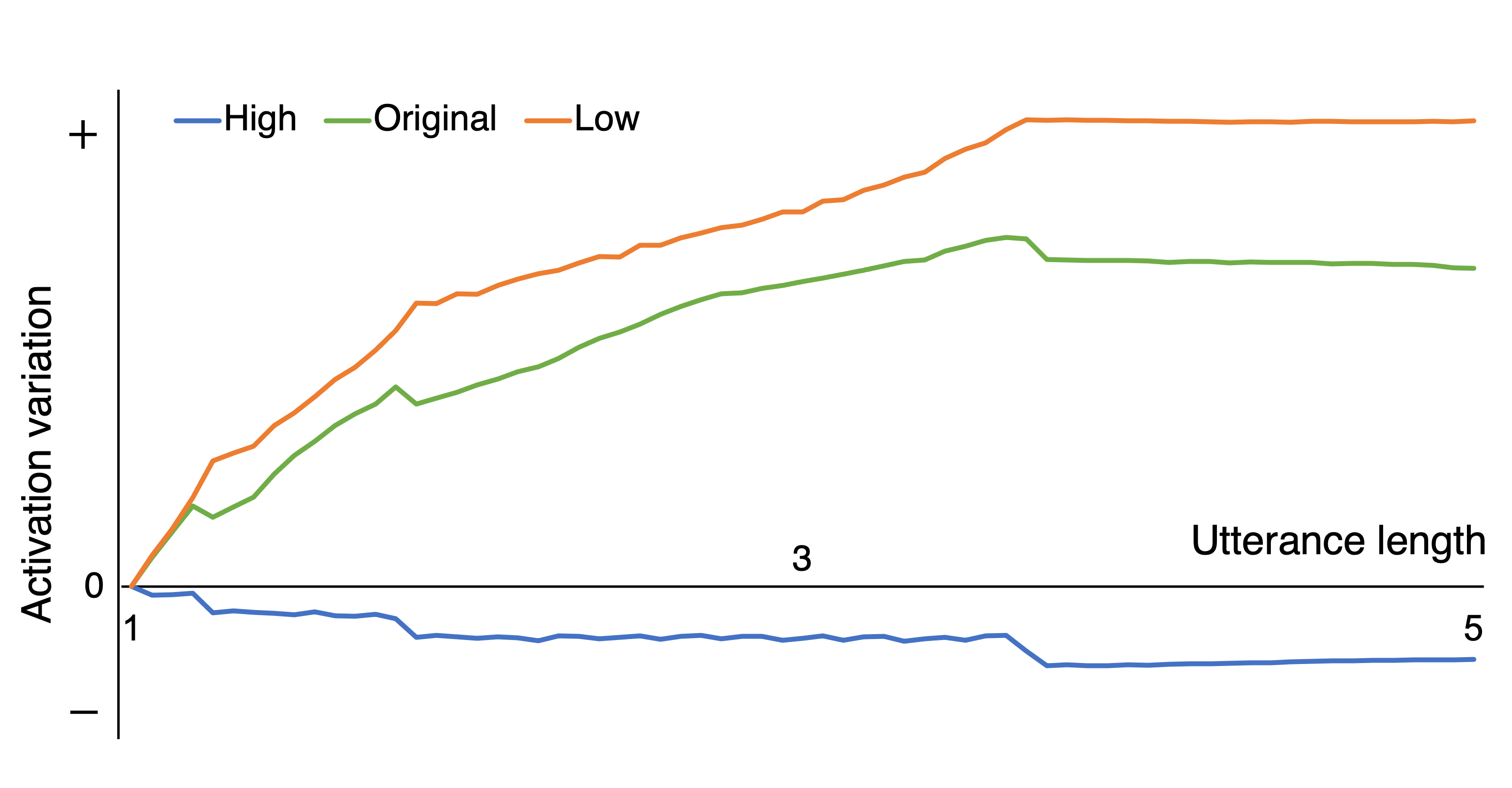}
  \caption{
  Variation score for mean activation of each resolution path according to the input utterance length on VoxCeleb1 test set. 
  }
  \label{fig:activation}
  \vspace{-0.2cm}
\end{figure}

Furthermore, to analyze the trained EDSP of RawNeXt, we defined the score $S^r_L$ at each $r$ resolution branch by differences of mean activations between $L$ and a 1-second utterance as follows: 
\begin{equation}
    S^{r}_L = \frac{1}{TC}(\sum_{t=1}^{T}\sum_{c=1}^{C}x^r_{L_{tc}} - \sum_{t=1}^{T}\sum_{c=1}^{C}x^r_{1_{tc}})
\end{equation}
where, $T$ and $C$ are the frame length and the number of channels of the feature, respectively. 
$x^r_L$ is the tensor of $L$ second utterance derived by multiplying the attention map with the activation of the $r$ resolution branch, as in each term of eq. (\ref{eq1}). 
Fig. \ref{fig:activation} visualizes the activation variation score (average over all layers) of each resolution branch according to the utterance length. 
Obviously, as the length of the input utterance increases, it becomes less active in the high-resolution path and more active in the low-resolution path. 
This tendency implies that the model can extract speaker information with appropriate resolutions by dynamically applying scaling policies according to the length of the utterance, as discussed in section \ref{subsec:edsp}. 
Thus, short utterances containing relatively sparse speaker information require exquisite feature extraction with small receptive fields at higher resolutions, whereas long utterances require more comprehensive feature extraction with large receptive fields at lower resolutions.

\section{Conclusion}
\label{sec:conclusion}
We proposed a novel speaker verification (SV) system, RawNeXt, using a deep layer aggregation (DLA) structure and extended dynamic scaling polices (EDSP) for variable-duration input utterances. 
The DLA extracts speaker embeddings rich in time-spectral information by aggregating the time scales and spectral channels of features in the network, iteratively and hierarchically. 
The EDSP dynamically captures speaker-discriminative information by selectively activating the branches of different resolutions in blocks according to the length of utterances. 
As a result of the evaluation on the VoxCeleb dataset, RawNeXt outperformed the recently proposed baseline systems for shorter utterances and demonstrated a strong generalization ability of SV while exhibiting comparable performance to the state-of-the-art systems. 
In addition, we proved the effectiveness of our system's components through ablation and analysis experiments.

\bibliographystyle{IEEEbib}
\bibliography{refs}

\begin{thebibliography}{10}

\bibitem{variani2014deep}
Ehsan Variani, Xin Lei, Erik McDermott, Ignacio~Lopez Moreno, and Javier
  Gonzalez-Dominguez,
\newblock ``Deep neural networks for small footprint text-dependent speaker
  verification,''
\newblock in {\em 2014 IEEE international conference on acoustics, speech and
  signal processing (ICASSP)}. IEEE, 2014, pp. 4052--4056.

\bibitem{snyder2018x}
David Snyder, Daniel Garcia-Romero, Gregory Sell, Daniel Povey, and Sanjeev
  Khudanpur,
\newblock ``X-vectors: Robust dnn embeddings for speaker recognition,''
\newblock in {\em 2018 IEEE International Conference on Acoustics, Speech and
  Signal Processing (ICASSP)}. IEEE, 2018, pp. 5329--5333.

\bibitem{dehak2010front}
Najim Dehak, Patrick~J Kenny, R{\'e}da Dehak, Pierre Dumouchel, and Pierre
  Ouellet,
\newblock ``Front-end factor analysis for speaker verification,''
\newblock {\em IEEE Transactions on Audio, Speech, and Language Processing},
  vol. 19, no. 4, pp. 788--798, 2010.

\bibitem{zhang2017end}
Chunlei Zhang and Kazuhito Koishida,
\newblock ``End-to-end text-independent speaker verification with triplet loss
  on short utterances.,''
\newblock in {\em Interspeech}, 2017, pp. 1487--1491.

\bibitem{jung2019short}
Jee-weon Jung, Hee-Soo Heo, Hye-jin Shim, and Ha-Jin Yu,
\newblock ``Short utterance compensation in speaker verification via
  cosine-based teacher-student learning of speaker embeddings,''
\newblock in {\em 2019 IEEE Automatic Speech Recognition and Understanding
  Workshop (ASRU)}. IEEE, 2019, pp. 335--341.

\bibitem{hajavi2019deep}
Amirhossein Hajavi and Ali Etemad,
\newblock ``A deep neural network for short-segment speaker recognition,''
\newblock {\em arXiv preprint arXiv:1907.10420}, 2019.

\bibitem{gao2019improving}
Zhifu Gao, Yan Song, Ian~Vince McLoughlin, Pengcheng Li, Yiheng Jiang, and
  Li-Rong Dai,
\newblock ``Improving aggregation and loss function for better embedding
  learning in end-to-end speaker verification system.,''
\newblock in {\em INTERSPEECH}, 2019, pp. 361--365.

\bibitem{jung2020improving}
Youngmoon Jung, Seong~Min Kye, Yeunju Choi, Myunghun Jung, and Hoirin Kim,
\newblock ``Improving multi-scale aggregation using feature pyramid module for
  robust speaker verification of variable-duration utterances,''
\newblock {\em arXiv preprint arXiv:2004.03194}, 2020.

\bibitem{Yu_2018_CVPR}
Fisher Yu, Dequan Wang, Evan Shelhamer, and Trevor Darrell,
\newblock ``Deep layer aggregation,''
\newblock in {\em Proceedings of the IEEE Conference on Computer Vision and
  Pattern Recognition (CVPR)}, June 2018.

\bibitem{ravanelli2018sincnet}
Mirco Ravanelli and Yoshua Bengio,
\newblock ``Speaker recognition from raw waveform with sincnet,''
\newblock in {\em 2018 IEEE Spoken Language Technology Workshop (SLT)}, 2018,
  pp. 1021--1028.

\bibitem{zhang2020deep}
Peng Zhang, Peng Hu, and Xueliang Zhang,
\newblock ``Deep embedding learning for text-dependent speaker verification.,''
\newblock in {\em INTERSPEECH}, 2020, pp. 3461--3465.

\bibitem{wang2019elastic}
Huiyu Wang, Aniruddha Kembhavi, Ali Farhadi, Alan~L Yuille, and Mohammad
  Rastegari,
\newblock ``Elastic: Improving cnns with dynamic scaling policies,''
\newblock in {\em Proceedings of the IEEE/CVF Conference on Computer Vision and
  Pattern Recognition}, 2019, pp. 2258--2267.

\bibitem{jung2020improved}
Jee-weon Jung, Seung-bin Kim, Hye-jin Shim, Ju-ho Kim, and Ha-Jin Yu,
\newblock ``Improved rawnet with feature map scaling for text-independent
  speaker verification using raw waveforms,''
\newblock {\em Proc. Interspeech 2020}, pp. 3583--3587, 2020.

\bibitem{nagrani2017voxceleb}
Arsha Nagrani, Joon~Son Chung, and Andrew Zisserman,
\newblock ``Voxceleb: a large-scale speaker identification dataset,''
\newblock {\em arXiv preprint arXiv:1706.08612}, 2017.

\bibitem{chung2018voxceleb2}
Joon~Son Chung, Arsha Nagrani, and Andrew Zisserman,
\newblock ``Voxceleb2: Deep speaker recognition,''
\newblock {\em arXiv preprint arXiv:1806.05622}, 2018.

\bibitem{nagrani2020voxsrc}
Arsha Nagrani, Joon~Son Chung, Jaesung Huh, Andrew Brown, Ernesto Coto, Weidi
  Xie, Mitchell McLaren, Douglas~A Reynolds, and Andrew Zisserman,
\newblock ``Voxsrc 2020: The second voxceleb speaker recognition challenge,''
\newblock {\em arXiv preprint arXiv:2012.06867}, 2020.

\bibitem{xie2017aggregated}
Saining Xie, Ross Girshick, Piotr Doll{\'a}r, Zhuowen Tu, and Kaiming He,
\newblock ``Aggregated residual transformations for deep neural networks,''
\newblock in {\em Proceedings of the IEEE conference on computer vision and
  pattern recognition}, 2017, pp. 1492--1500.

\bibitem{zhou2021resnext}
Tianyan Zhou, Yong Zhao, and Jian Wu,
\newblock ``Resnext and res2net structures for speaker verification,''
\newblock in {\em 2021 IEEE Spoken Language Technology Workshop (SLT)}. IEEE,
  2021, pp. 301--307.

\bibitem{okabe2018attentive}
Koji Okabe, Takafumi Koshinaka, and Koichi Shinoda,
\newblock ``Attentive statistics pooling for deep speaker embedding,''
\newblock {\em arXiv preprint arXiv:1803.10963}, 2018.

\bibitem{kye2021supervised}
Seong~Min Kye, Joon~Son Chung, and Hoirin Kim,
\newblock ``Supervised attention for speaker recognition,''
\newblock in {\em 2021 IEEE Spoken Language Technology Workshop (SLT)}. IEEE,
  2021, pp. 286--293.

\bibitem{kye2020meta}
Seong~Min Kye, Youngmoon Jung, Hae~Beom Lee, Sung~Ju Hwang, and Hoirin Kim,
\newblock ``Meta-learning for short utterance speaker recognition with
  imbalance length pairs,''
\newblock {\em arXiv preprint arXiv:2004.02863}, 2020.

\bibitem{lin2017feature}
Tsung-Yi Lin, Piotr Doll{\'a}r, Ross Girshick, Kaiming He, Bharath Hariharan,
  and Serge Belongie,
\newblock ``Feature pyramid networks for object detection,''
\newblock in {\em Proceedings of the IEEE conference on computer vision and
  pattern recognition}, 2017, pp. 2117--2125.

\bibitem{Tang2019DeepSE}
Yun Tang, Guo-Hong Ding, Jing Huang, Xiaodong He, and Bowen Zhou,
\newblock ``Deep speaker embedding learning with multi-level pooling for
  text-independent speaker verification,''
\newblock {\em 2019 IEEE International Conference on Acoustics, Speech and
  Signal Processing (ICASSP)}, pp. 6116--6120, 2019.

\bibitem{9292935}
Haiying Zhao, Wei Zhou, Xiaogang Hou, and Hui Zhu,
\newblock ``Double attention for multi-label image classification,''
\newblock {\em IEEE Access}, vol. 8, pp. 225539--225550, 2020.

\bibitem{torgashov2020id}
Nikita Torgashov,
\newblock ``Id r\&d system description to voxceleb speaker recognition
  challenge 2020,'' 2020.

\bibitem{xiang2020xx205}
Xu~Xiang,
\newblock ``The xx205 system for the voxceleb speaker recognition challenge
  2020,''
\newblock {\em arXiv preprint arXiv:2011.00200}, 2020.

\bibitem{thienpondt2021idlab}
Jenthe Thienpondt, Brecht Desplanques, and Kris Demuynck,
\newblock ``The idlab voxsrc-20 submission: Large margin fine-tuning and
  quality-aware score calibration in dnn based speaker verification,''
\newblock in {\em 2021 IEEE International Conference on Acoustics, Speech and
  Signal Processing (ICASSP)}. IEEE, 2021, pp. 5814--5818.

\bibitem{reddi2019convergence}
Sashank~J Reddi, Satyen Kale, and Sanjiv Kumar,
\newblock ``On the convergence of adam and beyond,''
\newblock {\em arXiv preprint arXiv:1904.09237}, 2019.

\bibitem{snyder2015musan}
David Snyder, Guoguo Chen, and Daniel Povey,
\newblock ``Musan: A music, speech, and noise corpus,''
\newblock {\em arXiv preprint arXiv:1510.08484}, 2015.

\bibitem{deng2019arcface}
Jiankang Deng, Jia Guo, Niannan Xue, and Stefanos Zafeiriou,
\newblock ``Arcface: Additive angular margin loss for deep face recognition,''
\newblock in {\em Proceedings of the IEEE/CVF Conference on Computer Vision and
  Pattern Recognition}, 2019, pp. 4690--4699.

\bibitem{chung2020defence}
Joon~Son Chung, Jaesung Huh, Seongkyu Mun, Minjae Lee, Hee~Soo Heo, Soyeon
  Choe, Chiheon Ham, Sunghwan Jung, Bong-Jin Lee, and Icksang Han,
\newblock ``In defence of metric learning for speaker recognition,''
\newblock {\em arXiv preprint arXiv:2003.11982}, 2020.

\end{thebibliography}

\end{document}